\documentclass[prl,aps,10pt,amsmath,amssymb,twocolumn,superscriptaddress]{revtex4-2}
\usepackage{amsmath,amssymb}
\usepackage[usenames]{color}
\usepackage{amssymb}
\usepackage{amsmath, amstext, amssymb, amsfonts, amsxtra}
\usepackage{textcomp}
\usepackage{xspace}
\usepackage{color}
\usepackage{float}
\usepackage{tikz}
\usepackage{xcolor}

\usepackage{graphicx}
\usepackage{hyperref}
\hypersetup{
   colorlinks = true,
   linkcolor=blue,
   citecolor=blue,
   urlcolor=blue
 }

\begin{document}
\title{Geometric filtering effect in expanding Bose-Einstein condensate shells}
\author{A. Tononi}
\affiliation{Department de F\'isica, Universitat Polit\`ecnica de Catalunya, Campus Nord B4-B5, E-08034, Barcelona, Spain}
\affiliation{ICFO-Institut de Ciencies Fotoniques, The Barcelona Institute of Science and Technology, Castelldefels (Barcelona) 08860, Spain}
\author{M. Lewenstein}
\affiliation{ICFO-Institut de Ciencies Fotoniques, The Barcelona Institute of Science and Technology, Castelldefels (Barcelona) 08860, Spain}
\affiliation{ICREA, Pg. Lluís Companys 23, 08010 Barcelona, Spain}
\author{L. Santos}
\affiliation{Institut f\"ur Theoretische Physik, Leibniz Universit\"at Hannover, Appelstrasse 2, 30167 Hannover, Germany}

\begin{abstract}
A shell-shaped Bose-Einstein condensate released from its confinement expands radially both outward and inward, displaying a self-interference pattern characterized by a density peak surrounded by a halo.
Here we analyze how an external imprinting or the thermal fluctuations of the condensate phase influence this expansion.
In both cases we find that the curved geometry filters the imploding finite angular-momentum modes via a radial centrifugal potential, so that only the condensate state can reach the origin and form the central peak.
As a consequence, we observe a pronounced dependence of the central density on the imprinting strength and on temperature.
This geometric filtering effect characterizes the free expansion of curved atomic gases in contrast with flat counterparts, it is easily observable in the available experimental platforms and enables two-dimensional shells thermometry via simple absorption-imaging techniques.
\end{abstract} 

\maketitle

Ultracold atom experiments can now confine quantum gases in spatially curved geometries \cite{carollo2021, jia2022, huang2025, dubessy2025, fernholz2007}, enabling the direct exploration of quantum physics in curved space \cite{tononi2023, moller2020}. Shell-shaped gases are a notable example \cite{tononi2024}, realized by magnetically \cite{carollo2021, lundblad2019} or optically \cite{jia2022, huang2025} trapping bosons near hollow surfaces such as spheres or ellipsoids.
A central question is how these shells expand once the confinement is switched off. Previous studies show that a pure condensate shell expands both inward and outward, forming a dense central peak --- a simple yet striking manifestation of the underlying curved geometry \cite{lannert2007, tononi2020, rhyno2021, elbourn2022, boegel2023}. For an initially pure condensate, the width of this peak is primarily determined by the strength of repulsive interatomic interactions \cite{lannert2007}. In contrast, the expansion of shells carrying nonzero angular momentum remains unexplored.

The problem is compelling because it parallels single-particle quantum mechanical motion in a spherically symmetric potential \cite{landau}: the angular momentum $l$ generates a centrifugal barrier $\propto l(l+1)/r^2$, repelling finite-$l$ modes from the origin. The interplay between radial motion and angular momentum underlies fundamental physical problems ranging from the hydrogen atom \cite{landau} to scattering \cite{landau, BransdenJoachain}, Efimov physics \cite{naidon2017, greene2017}, and nuclear reactions \cite{hippel1972}. The free expansion of shell-shaped gases at finite $l$ thus provides the opportunity to observe the centrifugal dynamics in a many-body quantum system.

In this Letter, we investigate the free expansion of condensate shells prepared with an angular-dependent phase profile. Such a phase can be imprinted externally or arise from thermally populated modes following a Bose-Einstein distribution. In both cases, as the shell expands toward the origin, the components with angular momentum $l>0$ experience the repulsive centrifugal potential $V_l(r) \propto l(l+1)/(2r^2)$. Consequently, while the $l=0$ condensate forms a central peak, higher-$l$ modes either bounce at a characteristic radial distance $r_l \propto \sqrt{l(l+1)}$ or convert into $l=0$ modes through the annihilation of the pair $(l,m)$, $(l,-m)$ into the $(l=0, m=0)$ state, where $m=-l,...,l$ denotes the angular momentum projection along the $z$ axis.

The central density thus depends on the initial phase or on the thermal population, providing in the latter case a {thermometry method} for two-dimensional (2D) shells. Our results, obtained for realistic experimental parameters \cite{jia2022, huang2025}, are directly observable in heteronuclear bosonic mixture experiments \cite{derrico2019, cavicchioli2025} and reveal a dynamical geometric filtering effect intrinsic to curved quantum gases.

\paragraph{Model.--} We consider a system of $N$ bosons of mass $M$ confined in a harmonic spherical-surface trap of radius $R$ and frequency $\omega_0$, given by the potential $U(r)=\frac{1}{2}M\omega_0^2 (r-R)^2$. 
The system is described by the grand-canonical Hamiltonian
\begin{equation}
\!\!\!\hat H =\!\!\int\!\!d^3 r \hat\Psi^\dag(\mathbf{r}) \!\left [ \hat H_r\! +\! \frac{\hat L^2}{2Mr^2}\!+\!\frac{g}{2}\hat\Psi^\dag(\mathbf{r})\hat\Psi(\mathbf{r})\! -\! \mu \right ]\!\hat \Psi(\mathbf{r}),
\label{Ham3D}
\end{equation} 
where $\hat{\Psi}^{\dagger}(\mathbf{r})$ is the bosonic creation operator at position $\mathbf{r}$, $\hat H_r = -\frac{\hbar^2}{2M}\nabla_r^2+U(r)$ is the radial Hamiltonian, with $\nabla_r^2 = \partial_r^2+\frac{2}{r}\partial_r$, $\hat L^2$ is the squared angular momentum operator, and $g=4\pi\hbar^2 a/M$ is the interaction strength, with $a$ the $s$-wave scattering length and $\mu$ the chemical potential.
We assume in the following that $R\gg l_0$, with $l_0=\sqrt{\hbar/(M\omega_0)}$ the harmonic oscillator length, and we set $\hbar\omega_0$ much larger than any other energy scale so that the system, while trapped, remains in the ground state of $\hat H_r$:
$\phi_0(r)= e^{-(r-R)^2/(2l_0^2)} / (\pi^{1/4}\sqrt{l_0} R)$, with eigenenergy $\hbar\omega_0/2$. We may hence write $\hat\Psi(\mathbf{r})=\phi_0(r)\hat\psi(\Omega)$, where 
$\Omega$ indicates the angular coordinates 
$(\theta\in [0,\pi], \varphi \in [0,2\pi])$. 
Integrating Eq.~\eqref{Ham3D} radially, we then obtain the 
Hamiltonian on the spherical surface:
\begin{equation}
\!\!\!\hat{H}_{\text{2D}}\! =\!\! \int\! d\Omega \hat\psi^\dag(\Omega) \!\!
\left [
\frac{\hat L^2}{2MR^2}\!+\!\frac{g_{2D}}{2} \hat\psi^\dag(\Omega) 
\hat\psi(\Omega)\! -\! \gamma \right ]\!\!\hat\psi(\Omega)
\label{Ham2D}
\end{equation}
where $\gamma = \mu-\hbar\omega_0/2$ is the chemical potential removing the radial energy, and 
$g_{\text{2D}} \equiv \frac{g}{\sqrt{2\pi}l_0 R^2}$. 
At zero temperature, the mean-field ground state is given by the uniform solution $\psi=\sqrt{n_0}$,
with $n_0 = N/(4\pi)$, characterized by the chemical potential $\gamma=g_{\text{2D}} n_0$ and having zero angular momentum. 
Note that the radial coordinate remains frozen in the trap as long as $\gamma \ll \hbar\omega_0$, which demands $\frac{Nal_0}{\sqrt{2\pi}R^2} \ll 1$. 
The results of this Letter are scaled in units of the length $R$, energy $E_R=\hbar^2/MR^2$, temperature $E_R/k_B$~(with $k_B$ the Boltzmann constant), and time $t_R=\hbar/E_R$.

\paragraph{Expansion of phase-imprinted states.--} 
The stationary mean-field trap state $\Psi_0(r) = \phi_0(r) \sqrt{n_0}$ has zero angular momentum.
Its expansion dynamics following the shell-trap removal [$U(r) = 0$] is already understood \cite{lannert2007}.
Qualitatively, the gas expands radially away from $r\sim R$, forming a dense peak at around $r=0$, surrounded by a broad halo. 
The peak density and width vary during the expansion, but they are essentially regulated by the interplay of the radial kinetic energy and of the interatomic repulsive interactions. 

The situation is much richer for initial states with nonzero angular momentum.
Such states can be experimentally prepared via a standard phase-imprinting protocol, consisting in applying for a time $\Delta t$ an external potential $V(\mathbf{r}) = v (x^2 + y^2 + \lambda^2 z^2)$, of strength $v$ and anisotropy $\lambda$, on the stationary condensate state $\Psi_0(r)$.
The initial phase-imprinted state reads $\Psi(\mathbf{r},t=0) = \phi_0(r) \sqrt{n_0} e^{i \Phi(\mathbf{r})}$, which is not any more spherically symmetric because it acquired the phase $\Phi(\mathbf{r}) = V(\mathbf{r}) \Delta t /\hbar$. 
If we express $V(\mathbf{r})$ in spherical coordinates around $r \sim R$, we get $\Phi = \Phi_0 + C Y_{20} (\Omega)$, where $\Phi_0$ is a dynamically irrelevant constant, $Y_{lm} (\Omega)$ are spherical harmonics, and $C = \sqrt{16\pi/45} (\lambda^2-1) (v R^2 \Delta t /\hbar)$ denotes the imprinting strength.
Therefore, the application of $V(\mathbf{r})$ just before the expansion creates a nonuniform phase field which, as we show below, corresponds to populating even angular momentum states. 
Such states have larger populations for larger values of $C$, provided that $C \lesssim (\gamma/E_R)$ holds to avoid creating significant density modulations. 


\begin{figure}[t!]
\centering
\includegraphics[width=0.98\columnwidth]{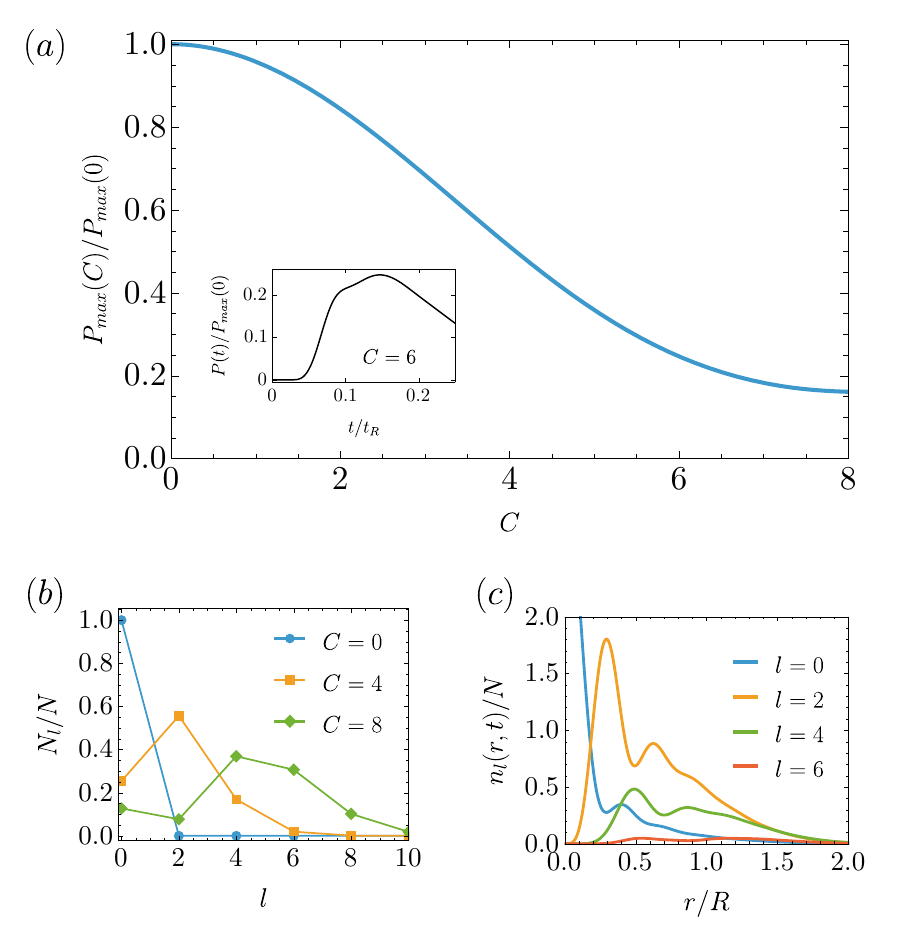}
\caption{(a): Maximal population ratio penetrating the radial region $r< R_{c}$ vs the phase-imprinting strength $C$.
When increasing $C$ in the imprinted state $\Psi(\mathbf{r},t=0)$, the initial occupation $N_l/N$ of even-$l$ angular momentum states shifts toward larger $l$ values [see (b)], and large-$l$ modes are more strongly repelled from the region $r=0$ by the centrifugal barrier [see (c)].
Here we choose the following realistic parameters \cite{jia2022}: $N=10^4$, $\gamma/E_R=7.7$, $l_0/R = 0.15$, and $R_{c}/R = 0.3$.
The densities of panel (c) are obtained for $C=6$ and are reported at the time at which the central population is maximal, $t=0.15 t_R$.
}
\label{fig1}
\end{figure}


We model the expansion dynamics of the initial state $\Psi(\mathbf{r},t=0)$ by solving the three-dimensional Gross-Pitaevskii equation:
\begin{equation}
i\hbar\dot\Psi(\mathbf{r},t) \!= \!\left [ -\frac{\hbar^2\nabla_r^2}{2M} + \frac{\hat L^2}{2Mr^2} +g |\Psi(\mathbf{r},t)|^2  \right ]\Psi(\mathbf{r},t),
\label{3DGPE}
\end{equation}
and monitor in particular how $C$ affects the maximal population penetrating the region $R_{c}$, with $R_{c} < R$.
We define it as
\begin{equation}
P_{max} = \max_{t} \left[ P(t) \right], \ P(t) = \int_{0}^{R_{c}} dr \, r^2 \int d\Omega \, |\Psi(\mathbf{r},t)|^2,
\end{equation}
and plot it as a function of $C$ in Fig.~\ref{fig1}(a). 
The maximal central population reached during the expansion decreases with stronger imprinting $C$, suggesting that, as larger angular momentum populations are imprinted onto the initial state, a smaller fraction of the gas is able to reach the central region.
This effect is qualitatively unchanged for different sizes of the internal region, here being $R_c = 0.3 R$.
To better visualize how the repulsion from the central region occurs, we calculate the initial imprinted populations $N_l = \int_0^{\infty} dr \, r^2 n_l (r,t=0)$, where the radial densities of the different angular-momentum modes are $n_l (r,t) = \sum_{m=-l}^l |\chi_{lm}(r,t)|^2$, and 
$\chi_{lm}(r,t) = \int d\Omega Y_{lm} (\Omega) \Psi(\mathbf{r},t)$. 
We report $N_l/N$ versus $l$ in Fig.~\ref{fig1}(b), which shows indeed that larger $C$ imprints a larger fraction of nonzero angular momenta onto the initial state.
Such finite-$l$ modes are, however, repelled from the origin, as the radial distributions $n_l(r,t)$ in Fig.~\ref{fig1}(c) demonstrate, and this filtering effect is responsible for the pronounced dependence on $C$ of the central population.
Note that the same repulsion effect, and qualitatively similar curves to those of Fig.~\ref{fig1}(a), can be obtained by analyzing the maximal central population of a thin condensate slice in the $xz$ plane, or of the column density integrated along $y$.

\paragraph{Geometric filtering.--} To gain analytical understanding on the expansion of the phase-imprinted state, we substitute the decomposition $\Psi(\mathbf{r},t) = Y_{00} \chi_{00}(r,t) + \sum_{lm}^{'} Y_{lm} (\Omega) \chi_{lm}(r,t)$ in Eq.~\eqref{3DGPE}, where $\sum^{'}_{lm} = \sum_{l\neq 0 m}$. 
After projecting over the spherical harmonics, we obtain
\begin{eqnarray}
i \hbar \dot{\chi}_{00} &=& -\frac{\hbar^2\nabla_r^2}{2M} \chi_{00} 
+ \frac{g}{4\pi} \, \bigg[ \big( |\chi_{00}|^2  + 2 \sum_{l m}{}^{'}|\chi_{lm}|^2 \big) \chi_{00} \nonumber \\ &+& \sum_{l m}{}^{'} (-1)^{m}\chi_{lm}\chi_{l-m} \, \chi_{00}^{*} \bigg], 
\label{componentsGPE-1}
\\
i \hbar \dot{\chi}_{l\neq 0 \, m}\! &=& \left[-\frac{\hbar^2\nabla_r^2}{2M} + V_l(r) \right] \chi_{lm} +
\frac{g}{4\pi} \, \big[ 2 |\chi_{00}|^2 \chi_{lm} \nonumber
\\
&+& (-1)^m  \chi_{00}^2 \chi_{l-m}^{*} 
\big], 
\label{componentsGPE-2}
\end{eqnarray}
where $V_l(r) = \hbar^2l(l+1)/(2Mr^2)$ is the centrifugal potential, 
and, for simplicity, we only report the terms with lower powers of the components $\chi_{l\neq 0 m}$ (i.e., we assume that $|\chi_{l\neq 0 m}| \ll |\chi_{00}|$).
At $t=0$, the amplitudes $\chi_{lm}(r,t=0)$ are Gaussian wave packets of width $l_0$ localized at around $r=1$. Their initial kinetic energy is proportional to $(4 l_0^2)^{-1}$ and it has variance $\sim l_0^{-2}$.

If, as a first approximation, we neglect the interactions by setting {$g = 0$}, the solution of the resulting linear equations shows that the condensate component $\chi_{00}$ broadens radially up to $r=0$, while the excited modes $\chi_{lm}$ face the centrifugal barrier $V_l(r)$, eventually vanishing beyond a turning point at a finite radius $r_l$. This radius can be evaluated by equating the initial kinetic energy of the wave packet $\propto (4 l_0^2)^{-1}$ with the potential energy $V_l(r_l)$, getting $r_l \sim l_0 \sqrt{l(l+1)}$. 
This means that the geometry filters high-$l$ modes, a fact which is clearly illustrated in Fig.~\ref{fig1}(c), which presents the even-$l$ components at the time at which the maximal central population is reached.

\begin{figure}[hbtp]
\centering
\includegraphics[width=0.98\columnwidth]{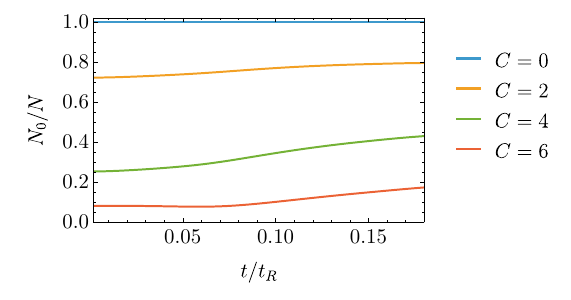}
\caption{Time evolution of $N_0/N$, demonstrating the role of nonlinear interactions during the inward implosion of the condensate shell. 
Note how, due the coupling at the second line of Eq.~\eqref{componentsGPE-1}, the population tends to convert into the condensate as the shell expands and the central density increases. 
We use for this figure the same parameter and scales of Fig.~\ref{fig1}.
}
\label{fig2}
\end{figure}

Let us now discuss qualitatively the role of the interactions by contrasting in particular the expansion of shells with the one of flat low-dimensional condensates \cite{dettmer2001, richard2003, hellweg2001}.
In the expansion dynamics of an elongated condensate~\cite{dettmer2001}, or the outward expansion of a shell condensate, the density quickly sinks. Interactions hence become rapidly negligible, and the evolution of the different momenta, or in the case of the shell-condensate angular momenta, may be considered independently. The situation is remarkably different in the inward expansion of the shell condensate, where the density increases due to the focusing effect associated to the spherical geometry. As a result, interactions play a relevant role in the inward expansion. 
A major consequence of this is the interconversion between the $l=0$ mode and the $l>0$ ones~[see the {$g$-dependent terms in Eqs.~\eqref{componentsGPE-1} and \eqref{componentsGPE-2}}]. 
This effect is clearly demonstrated in Fig.~\ref{fig2}, reporting $N_0/N$ versus time. 
Interestingly, also shells with relatively small initial populations of the $l=0$ mode dynamically increase their population thanks to the nonlinear interactions. 
Also note that the conversion between the excited modes and the condensate is initially slow and becomes faster as the central density peak forms.


Finally, let us compare the free expansion of phase-imprinted condensate shells and of stirred ring-shaped gases \cite{murray2013}. 
Ring-shaped Bose-Einstein condensates with initial angular momentum $m$, the eigenvalue of the angular momentum projection along $z$, $\hat{L}_z$, display a central hole with radius increasing linearly with $m$ \cite{murray2013}. 
This effect is likely due to an \textit{axial} centrifugal potential $\hbar^2 m^2/[2M (x^2+y^2)]
$ repelling high-angular-momentum modes.
This repulsion is analogous to what we observe in the shell-expansion case, although the centrifugal potential $V_l(r)$ has \textit{spherical} rather than axial symmetry, and the angular momentum modes are more strongly repelled due to the potential scaling as $l^2+l$ instead of $m^2$.

\begin{figure}[hbtp]
\centering
\includegraphics[width=0.98\columnwidth]{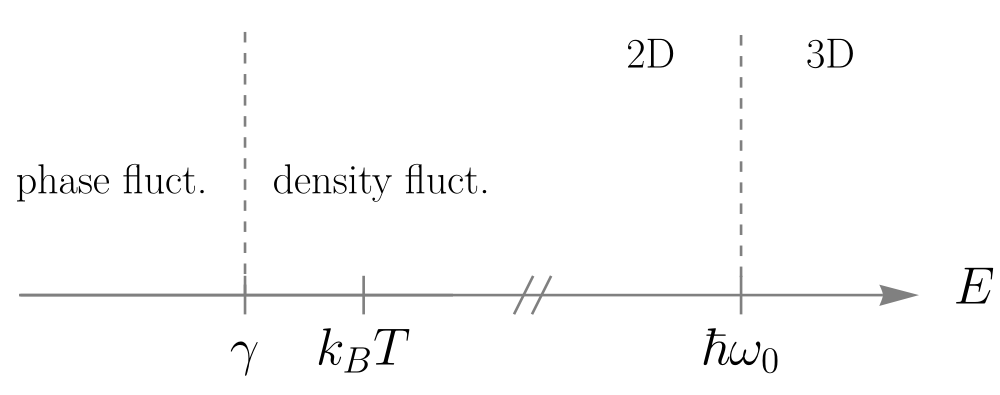}
\caption{Hierarchy of energy scales identifying different regimes of trapped bosonic gases.
We implement a 2D theory that includes both phase and density fluctuations to describe the free expansion of a two-dimensional shell-shaped Bose-Einstein condensate in the regime $ \gamma \lesssim k_B T \ll \hbar\omega_0$.
}
\label{fig3}
\end{figure}

\paragraph{Finite-temperature expansion.--} 
Finite-temperature condensate excitations provide another mechanism to get initial states with nonzero populations of the angular momentum components, and therefore of observing the filtering effect.
We analyze below the expansion dynamics of 2D spherically symmetric shells in the low-temperature regime where both condensate phase and density fluctuations are thermally excited.
This corresponds to the regime $\gamma \sim k_B T \ll \hbar \omega_0$, with $k_B$ the Boltzmann constant, as depicted in Fig.~\ref{fig3}.


Let us derive the phase and density fields by implementing the Bogoliubov theory. 
First, we expand the field operator in Eq.~\eqref{Ham2D} in the Madelung representation $\hat {\psi}(\Omega) = \sqrt{n_0+\hat{\delta n}(\Omega)} \, e^{i\hat{\Phi}(\Omega)}$, where $\hat{\delta n}(\Omega)$ and $\hat{\Phi}(\Omega)$ are, respectively, operators accounting for density and phase fluctuations around the mean-field uniform condensate solution $\psi$. 
From the linearization of the Heisenberg equation of $\hat\psi$ under $\hat H_{\text{2D}}$, we obtain the coupled 
equations $\hbar\partial_t \hat\Phi = -[\hat L^2/(2MR^2)+2\gamma](\hat{\delta n}/2n_0)$ and $\hbar\partial_t (\hat{\delta n}/2n_0) = [\hat L^2/(2MR^2)] \hat\Phi$.
We then diagonalize these equations via the Bogoliubov transformations $2\sqrt{n_0} \hat{\Phi}(\Omega) = \sum_{lm} \left [ f_{lm}^+ Y_{lm}(\Omega) e^{-i E_l t/\hbar} \hat a_{lm} + \text{H.c.}  \right ]$ and 
$\hat{\delta n}(\Omega)/\sqrt{n_0} =  \sum_{lm} \left [ i f_{lm}^- Y_{lm}(\Omega) e^{-i E_l t/\hbar} \hat a_{lm} + \text{H.c.} \right ]$, 
where $\hat{a}_{lm}$ are the destruction operators of bosonic quasiparticle excitations. Performing this diagonalization provides the excitation spectrum $E_l = \sqrt{\epsilon_l \left ( \epsilon_l + 2\gamma \right) }$, with $\epsilon_l = \hbar^2l(l+1)/(2MR^2)$ the single-particle energy, and the amplitudes $f_{lm}^{+}  = f_l^{+} = \sqrt{E_l/\epsilon_l}$ and $f_{lm}^- = 1/f_l^+$.

We model the free expansion of a finite-temperature shell-shaped initial state by taking the initial condition $\Psi(\mathbf{r},t=0) = \phi_0(r) \sqrt{n_0 + \delta n (\Omega)}  \, e^{i\Phi(\Omega)}$, 
where the phase and density fields $\Phi(\Omega)$ and $\delta n (\Omega)$, defined as
\begin{align}
\begin{split}
\Phi(\Omega) &= \frac{1}{\sqrt{n_0}} \sum_{l=1}^{\infty} \sum_{m=-l}^l  f_{l}^{+} \, \text{Re} [ Y_{lm}(\Omega) \alpha_{lm}],
\\
\delta n(\Omega) &= \sqrt{n_0} \sum_{l=1}^{\infty} \sum_{m=-l}^l f_{l}^-  \, \text{Re} [ i  Y_{lm}(\Omega) \alpha_{lm} ],
\label{phase}
\end{split}
\end{align}
are stochastic spatially dependent quantities distributed in accordance to $\hat{\Phi}(\Omega)$ and $\hat{\delta n}(\Omega)$.
We derive their expressions by replacing the operators $\hat a_{lm}$ in $\hat{\Phi}_{\Omega}$ and $\hat{\delta n}(\Omega)$ with the complex random variables $\alpha_{lm}$ with mean $\langle \alpha_{lm} \rangle = 0$ and variance $\langle |\alpha_{lm}|^2 \rangle = N_l^{B}(T)$, with $N_l^{B}(T) = 1/[e^{E_l/(k_B T)}-1]$ the Bose-Einstein distribution at temperature $T$.
To correctly account for thermal effects, 
we generate many stochastic realizations of the phase fields $\Phi(\Omega)$ and $\delta n(\Omega)$ by sampling the uniform random variables $\beta_{lm},\nu_{lm}\in[0,1]$ and calculating $\alpha_{lm} = (-N_l^{B}(T) \ln{\beta_{lm}})^{1/2} e^{i 2 \pi \nu_{lm}}$.
The finite-temperature expansion dynamics is modeled by solving for each phase and density realization the three-dimensional Gross-Pitaevskii equation \eqref{3DGPE} and averaging the observables over realizations. 

\begin{figure}[t!]
\centering
\includegraphics[width=0.98\columnwidth]{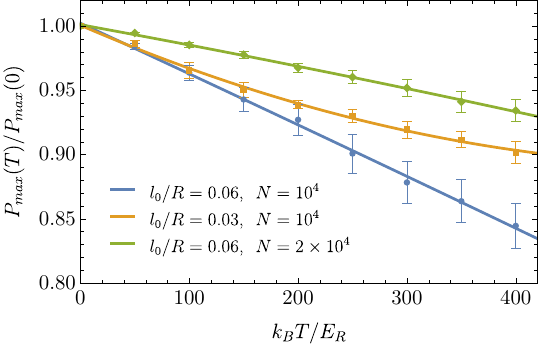}
\caption{Maximal population ratio penetrating the radial region $r< R_{c}$ vs temperature.
Similarly to the phase-imprinting case, larger temperatures produce condensate states with larger initial angular momenta, which are impeded to reach the central radial regions by the centrifugal potential repulsion. 
At a given temperature $T$, the maximal central population tends to decrease either when the shell thickens at fixed $N$, or when $N$ decreases at fixed shell thickness.
Note that we consider here a thin shell in the 2D regime of Fig.~\ref{fig3}, that the critical temperatures for $N = \{ 1 , 2 \} \times 10^4$ are $\{ 800, 1600\} \, k_B/E_R$, and we use for this figure $R_{c}/R = 0.3$.
The error bars show the standard deviation obtained from averaging over many realizations of the initial state.
}
\label{fig4}
\end{figure}

We plot the fractional maximal central population in Fig.~\ref{fig4} which, similarly to the phase-imprinting case, decreases with temperature. 
The effect is less pronounced with respect to Fig.~\ref{fig1}, essentially because the phase imprinting can transfer much larger populations to nonzero angular momentum modes with respect to the thermally induced case, that is limited by the Bose-Einstein distribution. 
Still, the central population decreases with temperature, demonstrating also in this case the dynamical filtering effect by the curved geometry.
We find that the central population ratio is affected by the parameters' choice, and that it tends to decrease with lower three-dimensional gas densities in the regimes explored.
The method presented here, therefore, enables calculating the maximal central density $P_{\text{max}}$ versus $T$ for given $R$, $l_0$, $N$, and $a$ in input.
Conversely, the temperature of spherical shells can be deduced by comparing the measured the central region density against the simulated one, which constitutes a method for spherical shells thermometry.
This effect provides, therefore, a tool for the experimental analyses of shell-shaped Bose-Einstein condensates.

\paragraph{Experimental relevance.--} 
The parameters of this Letter are chosen to be plausible for realistic implementations \cite{jia2022, huang2025, wolf2022, ma2025, veyron2025}.
Indeed, out of rescaling, Figs.~\ref{fig1} and \ref{fig2} parameters correspond to $10^4$ $^{23}\text{Na}$ atoms, forming a shell of radius $R=10 \, \mu\text{m}$ and thickness $l_0=1.5 \, \mu\text{m}$.
The two-dimensional rescaled interaction strength $\gamma/E_R = 7.7$ corresponds to a three-dimensional s-wave scattering length of $55$ Bohr radii, and the expansion times are always in the millisecond range. 
We take the same physical parameters in Fig.~\ref{fig4} except for the thicknesses $l_0=\{ 0.3, 0.6 \} \, \mu\text{m}$, and the maximum represented temperature $T = 400 \, E_R/k_B$ equals $84\,\text{nK}$, which is lower than the critical temperatures $\sim \{ 170, 340 \}\,\text{nK}$ for $N = \{ 1, 2  \}\times 10^4$  \cite{tononi2019}.
{These values are realistic and within reach of the experiments \cite{jia2022, huang2025}, where forthcoming improvements on atom numbers should allow controlling the shell thickness and manipulate it locally \cite{privcomm}.
In particular, Refs.~\cite{jia2022,huang2025} already} analyzed the expansion of three-dimensional (3D) condensate shells and, notably, employed the Abel transform to get radial densities from the measured column densities.
We mention that, in principle, the filtering effect described by Eqs.~\eqref{componentsGPE-1} and \eqref{componentsGPE-2} can be observed also in 3D shells, and compared with the Gross-Pitaevskii equation dynamics starting from a 3D ground state with an imprinted $\mathbf{r}$ dependent phase profile.
The advantage of taking initial 2D conditions is, however, that a tighter confinement produces initial wave packets with larger kinetic energy and narrower radial momentum distribution, making the filtering effect of imprinted shells more visible. 
In any case, we expect that Feshbach resonances in these or other heteronuclear mixtures experiments \cite{derrico2019, cavicchioli2025} will give good margins of tunability on the shell thickness.


\paragraph{Conclusions.--} We analyzed the free expansion of condensate shells with nonzero initial angular momentum state.
We find that the interplay between the system dynamics and the centrifugal potential dynamically purifies the initial state by filtering the expanding modes with finite angular momentum during the inward radial expansion.
Our simulations are realistic and close to the parameters of the experiments \cite{jia2022}, where the filtering effect can be directly observed.
Moreover, the simulation method introduced in this work can be combined with precise measurements of the central density of expanding spherical shells to precisely determine their temperature, enabling the observation of their finite-temperature phase diagram \cite{tononi2019}.
The observation of our predictions would constitute a direct experimental probe of the nonlinear dynamics of an interacting condensate in the radial centrifugal potential,
and it unequivocally demonstrates how spatial curvature affects the dynamics of ultracold atomic gases.
Future extensions of our work may elucidate how topological phase patterns such as vortex pairs and domain walls affect the time-of-flight expansion and interplay with the radial centrifugal potential.

\vspace*{5mm}
\begin{acknowledgements}
We thank N. Lundblad for useful discussions. A.T. acknowledges funding by the European Union under the Horizon Europe MSCA programme via the project 101146753 (QUANTIFLAC) and support by the Spanish Ministerio de Ciencia, Innovación y Universidades (grant PID2023-147469NB-C21, financed by MICIU/AEI/10.13039/501100011033 and FEDER-EU).
L.S. acknowledges the support of the Deutsche Forschungsgemeinschaft (DFG, German Research Foundation) under Germany's Excellence Strategy -- EXC-2123 Quantum-Frontiers -- 390837967.
ICFO-QOT group acknowledges support from:
European Research Council AdG NOQIA; MCIN/AEI [PGC2018-0910.13039/501100011033, CEX2019-000910-S/10.13039/501100011033, Plan National STAMEENA PID2022-139099NB, project funded by MCIN and by the “European Union NextGenerationEU/PRTR" (PRTR-C17.I1), FPI]; QUANTERA DYNAMITE PCI2022-132919; QuantERA II Programme co-funded by European Union’s Horizon 2020 program under Grant Agreement No 101017733; Ministry for Digital Transformation and of Civil Service of the Spanish Government through the QUANTUM ENIA project call - Quantum Spain project, and by the European Union through the Recovery, Transformation and Resilience Plan - NextGenerationEU within the framework of the Digital Spain 2026 Agenda; MICIU/AEI/10.13039/501100011033 and EU (PCI2025-163167); Fundació Cellex; 
Fundació Mir-Puig; Generalitat de Catalunya (European Social Fund FEDER and CERCA program; Barcelona Supercomputing Center MareNostrum (FI-2023-3-0024); 
Funded by the European Union (HORIZON-CL4-2022-QUANTUM-02-SGA, PASQuanS2.1, 101113690, EU Horizon 2020 FET-OPEN OPTOlogic, Grant No 899794, QU-ATTO, 101168628), EU Horizon Europe Program (No 101080086 NeQSTGrant Agreement 101080086 — NeQST).

\end{acknowledgements}

The data that support the findings of this article are openly available ~\cite{data}.

\end{document}